\documentclass[]{revtex4-1}
\usepackage{amsmath}    
\usepackage{graphicx}   
\usepackage{verbatim}   
\usepackage{color}      
\usepackage{subfigure}  
\usepackage{hyperref}   
\usepackage{lineno} 
\usepackage{bm} 
\usepackage{multirow}
\usepackage{csquotes}
\usepackage{appendix}
\usepackage{caption}
\usepackage{subcaption}
\newcommand{\classoption}[1]{\texttt{#1}}

\expandafter\ifx\csname package@font\endcsname\relax\else
\expandafter\expandafter
\expandafter\usepackage
\expandafter\expandafter
\expandafter{\csname package@font\endcsname}%
\fi
\DeclareRobustCommand\substyle{\name@idx{document substyle}}%
\DeclareRobustCommand\classoption{\name@idx{document class option}}%
\DeclareRobustCommand\classname{\name@idx{document class}}%
\def\name@idx#1#2{
	{\ttfamily#2}%
	\index{#2\space#1=\string\ttt{#2}\space#1}\index{#1>#2=\string\ttt{#2}}%
}

\begin{document}

	\pagenumbering{arabic}
	\title{A method for energy and radius reconstruction with simulated charge information in a ton-scale liquid scintillator detector like the Taishan Antineutrino Observatory}
\author{Randhir Singh$^{1,2,a}$,
	Yichen Li$^1$
	and
	Xiaochuan Xie$^3$}

\affiliation{Institute of High Energy Physics, Chinese Academy of Sciences, Beijing 100049, China$^1$\\
	Present Address: Department of Physical Sciences, Bose Institute, EN-80, Sector V, Kolkata 700091, India$^2$\\
	TIANFU Cosmic Ray Research Center, No.1500 Kezhi Road, Chengdu, China $^3$}
	\email{randhir.singh@ihep.ac.cn$^a$}

	\begin{abstract}
Small neutrino detectors are a ton level detectors which can be placed very close to the core of Nuclear Power Plant. In some detectors liquid scintilling (LS) material is used as the detecting material. The antineutrinos from the reactor core fall on the liquid scintillator of the detector where they deposit energy via Inverse Beta Decay process (IBD). The energy absorbed by liquid scintillator is re-emitted in the form of scintillation. These photons then travel through the scintillating material and hit the Silicon Photo Multipliers (SiPMs) which are installed on the inner surface of detector's spherical copper shell. These SiPMs absorb the photons to give a charge output signal. The energy and radius reconstruction is done using the information of charge collected by the SiPMs. Due to factors like large photo-coverage with large photon detection efficiency, small spherical detector size and low temperature operation, small size LS detectors can achieve an unprecedented energy resolution. In this paper, we have used a template-dependent method exploiting simulated data to reconstruct the event radius and energy. This method uses response functions generated using radioactive source calibration data and the charge information to reconstruct the energy and the radius of the event by constructing and maximizing a likelihood function. This methodology is applicable to all similar size spherical neutrino detector experiments.\\
 
	\end{abstract}
    \maketitle

\section{Introduction}
\label{intro}
Liquid scintillator (LS) detectors have been widely used in neutrino experiments as the detecting medium. Experiments such as Kamioka Liquid Scintillator Antineutrino Detector (KamLAND) \cite{KamLAND:2002uet}, Borexino \cite{Redchuk:2020igv}, Double Chooz \cite{DoubleChooz:2011ymz}, Daya Bay \cite{DayaBay:2012fng}, and Reactor Experiment for Neutrino Oscillation (RENO) \cite{RENO:2012mkc} have used liquid scintillator as the detection medium. Recently completed world's largest LS detector JUNO (Jiangmen Underground Neutrino Observatory) \cite{JUNO:2021vlw,JUNO:2025gmd} experiment is also a 20 kton liquid scintillator detector which has started taking data. The JUNO experiment achieves an approximate energy resolution of 3.5$\%$ for $^{68}Ge$ \cite{JUNO:2025gmd}, whereas experiments such as the Borexino and KamLAND achi-eved the energy resolution levels of $5\%$ \cite{BOREXINO:2026owb} and $6.5\%$ \cite{KamLAND:2002uet} at 1MeV, respectively. The JUNO detector has been designed to achieve high LS light yield, high transparency, and high photon collection efficiency so that the statistical fluctuation of the detected Photoelectrons (PE) can be minimized. JUNO will detect the antineutrinos via inverse beta decay (IBD) interactions, $\overline{\nu}_{e} + p \rightarrow e^{+} + n$. The prime goals of JUNO Experiment are to determine the Neutrino Mass Hierarchy \cite{Zhan:2008id} and the precise measurement of Neutrino Oscillation Parameters. In addition to that, it will also look for Neutrinoless Double Beta Decay \cite{Dolinski:2019nrj}, search for sterile neutrinos \cite{Gariazzo:2015rra,Berryman:2021xsi} and study of dark matter and astrophysical neutrinos\cite{JUNO:2023vyz}. The main source of antineutrinos for JUNO are the nuclear reactors at the Yangjiang Nuclear Power Plant and the Taishan Nuclear Power Plant. The small oscillation peaks in the oscillated antineutrino spectrum contain the Neutrino Mass Ordering (NMO) information. Therefore, the precise measurement of the oscillated antineutrino spectrum is important for JUNO to determine the NMO and the determination of the neutrino oscillation parameters.

Small neutrino detectors are a ton level detectors. These are small in size and are very compact (like JUNO-TAO \cite{JUNO:2020ijm}). Such a ton-level spherically shaped liquid scintillator (LS) detector can be placed very close to a reactor core of the a Nuclear Power Plant (NPP). Since experiments like Daya Bay, Double Chooz, RENO and others have found discrepancies between the data and the model prediction on the reactor antineutrino spectrum \cite{Huber:2011wv,Mueller:2011nm}, this type of small near detectors would provide a model-independent reference spectrum which will reduce the spectrum uncertainty. They also provide a benchmark measurement to test nuclear databases. Moreover, such a compact detector would be able to test the reactor antineutrino anomaly and search for light sterile neutrinos at the eV scale \cite{JUNO:2020ijm} and also has some potential through elastic neutrino-electron scattering (E$\nu$ES) \cite{Delgadillo:2026jee}. 

The reconstruction of the individual events is one of the most important task in order to obtain neutrino information such as their energies. In general, the event reconstruction should provide us the information about the energy, vertex and the interaction time (relative to the event trigger time) of the particle in the detecting material. The accuracy of the reconstruction depends on the properties of the event, the detector response, the performance of the the photon detectors and the electronic readout system.

Previous studies have reconstructed the vertex and the energy of the event by applying various techniques and methods. One of them being a likelihood fit which is based on an analytical optical model \cite{Wen:2011zzb}. In this method, the temporal and spatial distributions of photoelectrons (PEs) are precisely calculated by ignoring any scattering effect during photon's propagation. JUNO also uses the charge and time information to reconstruct the energy and the vertex of the events \cite{JUNO:2024fdc}. Preliminary results reveal that an energy resolution of 2.95$\%$ at 1 MeV can be achieved. The application of modern machine learning algorithms enabled the reconstruction of event vertex and energy in JUNO with a resolution of 10 cm and 3$\%$, respectively, at 1 MeV \cite{Qian:2021vnh}. In Ref. \cite{Shi:2025zut} a performance comparison of two vertex reconstruction techniques developed for these small experiments is presented: a traditional Charge Center Algorithm (CCA) and a novel Deep Learning Algorithm (DLA). Following dedicated optimization efforts, the CCA achieves a vertex resolution lower than 20 mm (bias lower than 5 mm), successfully meeting the experiment's requirements. The DLA significantly outperforms this, achieving a superior resolution of lower than 12 mm with a minimal bias lower than 1.3 mm at 1 MeV. A data driven calibration approach to reconstruct simultaneously the energy and the vertex is also explored \cite{Huang:2022zum}. A more realistic SiPM charge response model, that includes all relevant electronic effects, and a more accurate photon hit time probability distribution function (PDF) are constructed from Germanium-68 ($^{68}\text{Ge}$) calibration data instead of simulated positron data. The time PDF accounts for dependencies on both the event's vertex radius and the photon's propagation distance. The study in Ref \cite{Liu:2024cxo} applied a Tweedie generalized linear model (GLM), based on an inhomogeneous Poisson process framework, to model the SiPM's detector response and its output charge distribution. Based on time and charge of SiPMs, a pure probabilistic model was developed from first principles to reconstruct point-like events in the central detector of such a compact LS experiment and achieved a vertex resolution better than 20mm. The energy of the event can be also reconstructed based on the total number of collected photoelectrons based on calibration data, as done in Daya Bay \cite{DayaBay:2016ggj}. It would be useful to find a method to derive the charge responses and then reconstruct individual event energies from the calibration data using the spherical symmetry of the detector. In this paper, we have used a method which uses the templates generated with simulated calibration data to estimate energy and radius. 

The paper is organised as follows. In section \ref{Detector}, we will briefly discuss about detector setup. Section \ref{Algorithm} is dedicated to the introduction of the reconstruction met-hod. Section \ref{Results} presents the performance results and discussion, including performance analysis. Finally, Section \ref{conclusion} concludes the paper and outlines potential future work.

\begin{figure}[!h]
	\includegraphics[width=10cm,height=12cm]{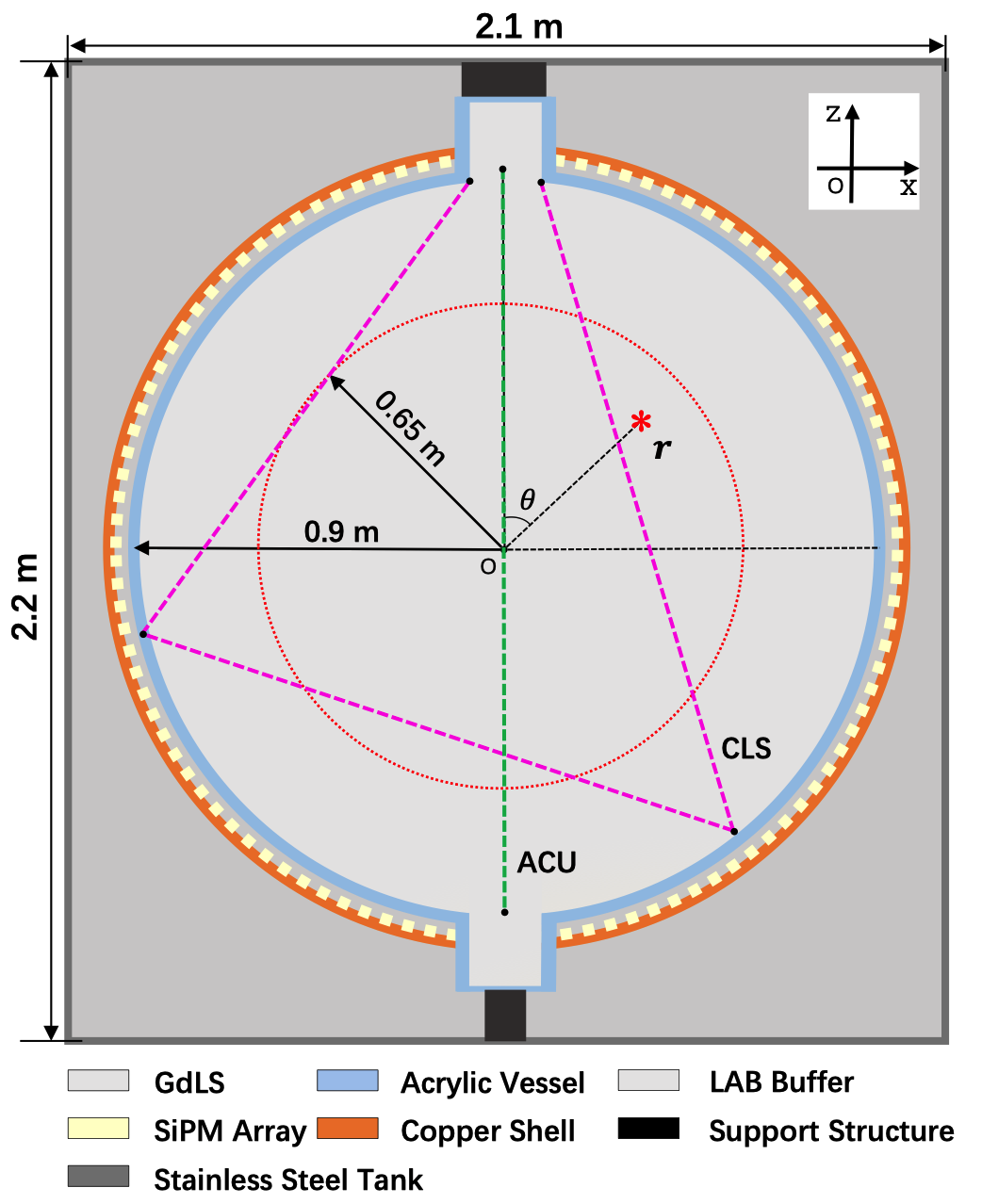}
	\caption{The detector setup, which consists of the central detector, the calibration system, consisting of an Automated Calibration Unit (ACU) and a Cable Loop System (CLS)\cite{Shi:2025zut}.}
	\label{detector-setup}
\end{figure}
\section{The Detector Setup}
\label{Detector}

Fig. \ref{detector-setup} shows the schematic of the detector setup. It has a Central Detector (CD), the calibration system, the outer shielding and the veto system. The CD consists of a spherical acrylic vessel, and it contains Gadolinium-doped Liquid Scintillator (GdLS). The LS mixture is based on Linear Alkylbenzene (LAB) because of its excellent transparency, high flash point, low chemical reactivity and good light yield. The motivation behind loading LS with 0.1$\%$ gadolinium is that the neutron capture by Gd will produce a delayed signal of about 8 MeV (much higher than natural radioactivity) which will effectively reject the accidental backgrounds. The LS also consists of small amount of about 3 g/L 2,5-diphenyloxazole (PPO) as the fluor and 2 mg/L p-bis-(o-methylstyryl)-benzene (bis-MSB) as the wavelength shifter. A small amount (0.5$\%$) of DPnB (Dipropylene glycol n-butyl ether) is added as cosolvent. The deposition of the energy ($E_{dep}$) of positrons primarily occurs through excitation and ionization processes. Due to very small concentrations of PPO and bis-MSB in the LS, the deposited energy is mainly transferred to LAB molecules. Subsequently, the excited LAB molecules can transfer their energy to fluor molecules through complex molecular-scale processes. When these fluor molecules are de-excited, scintillation photons are emitted. The acrylic vessel is placed inside a spherical copper shell. On the inner surface of the copper shell, Silicon Photo-multipliers (SiPMs) tiles are installed with a coverage of $\sim$10m$^2$. Each SiPM tile has two channels and has an area of about 25cm$^2$ and consists of
an 8x8 array of SiPMs with dimensions of 6×6mm$^2$. The SiPMs have a photon detection efficiency of $\sim$ 50$\%$ and a photo coverage larger than 95$\%$. To suppress the contribution of the dark noise of the SiPMs, the detector has to operate at -50 $^{o}$C which is maintained using a cryostat. The copper sphere not only provides the acrylic vessel with the mechanical support but also keeps the SiPM tiles pointing towards the center of the detector. On the outer surface of the copper sphere, cooling pipes and the readout electronics are installed. The central detector is placed in stainless steel water tank (SST) which contains the cryogenic part of the detector and also provides shielding from the environmental radioactivity from the rock and air. The water tanks have 3-inch Photomultiplier Tubes (PMTs) installed to detect the Cherenkov light from cosmic muons and will act as a veto detector. On top of the detector, layers of High Density Polyethylene (HDPE) are placed on which provide a passive shielding against the neutrons produced by the cosmic muons and the radioactivity from the materials outside the detector. The HDPE shielding has a cover of a plastic scintillator layer on its top for tagging the cosmic muons. At the bottom of the detector, lead bricks and HDPE panels are laid to provide a passive shield against the radioactivity. A complete picture of such a detector setup can be found in ref \cite{Xu:2022mdi}(fig.1).\\

Calibration is critical to achieve the goals for a high energy resolution, $<$1$\%$ uncertainty in the energy non-linearity and $<$0.5$\%$ residual non-uniformity. The calibration system consists of the Automated Calibration Unit (ACU) (green line in Fig.\ref{detector-setup}) and a Cable Loop System (CLS) (pink line in Fig.\ref{detector-setup}) . The ACU and CLS calibrate the energy response along the central axis (Z-axis) and off the central axis, respectively. A $^{68}\text{Ge}$ source, an ultraviolet (UV) light source (which is a UV LED calibration subsystem), and multiple gamma sources are installed in the ACU, whereas a $^{137}\text{Cs}$ source is deposited onto a small section of the CLS cable for calibration along off central axis \cite{Xu:2022mdi}.


\section{The Charge Likelihood Algorithm}
\label{Algorithm}

\begin{figure}[!h]
	\includegraphics
	[width= 9cm, height=10cm]
	{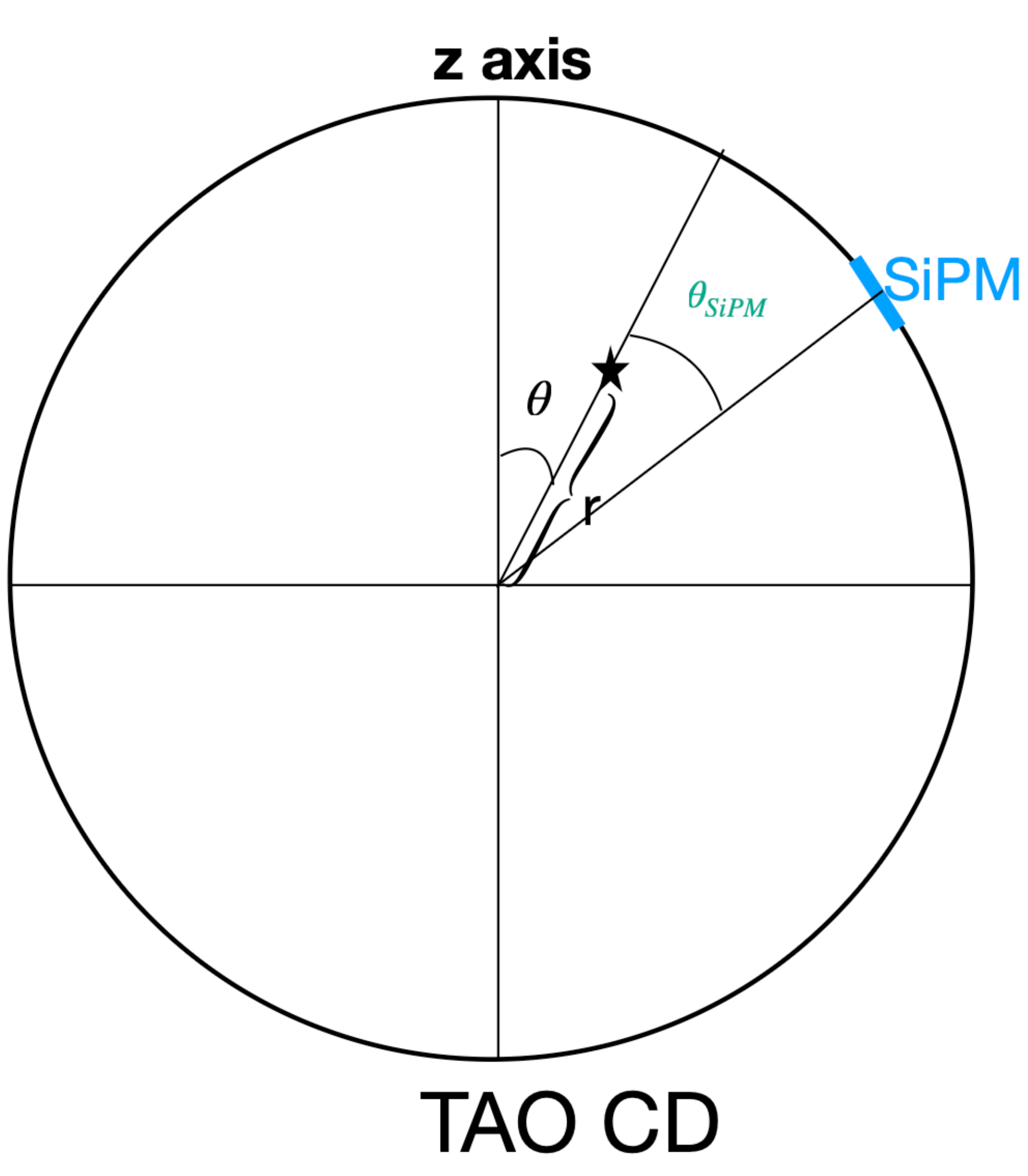}
	\caption{Event parameters in CD.}
	\label{event}
\end{figure}

\begin{figure}[!h]
	\includegraphics
	[width= 12cm, height=10cm]
	{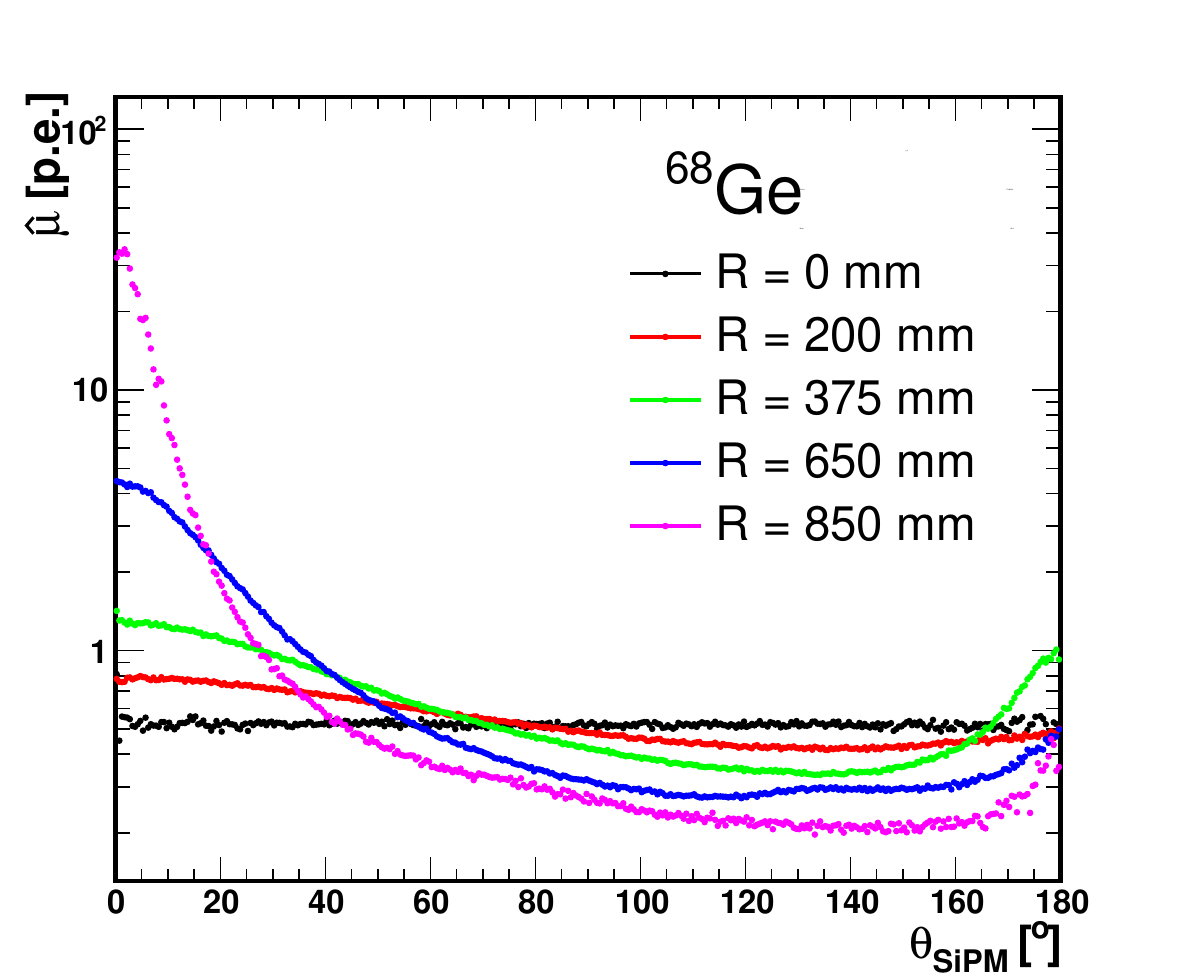}
	\caption{Simulated nPE maps generated by $^{68}\text{Ge}$ (2$*$0.511 MeV gamma source).}
	\label{templates}
\end{figure}

In this section, we discuss the charge likelihood algorithm which will rely on the templates generated with calibration data to reconstruct the energy and radius of IBD positrons in the CD. The charge likelihood algorithm is based on the distribution of the number of photoelectrons (nPE) in each SiPM  \cite{Wu:2018zwk}. The nPE detected by SiPMs are contributed by the absolute light yield, the attenuation of propagation mediums, the geometry effect and the detection efficiency of SiPMs. The main part of this method is the expected nPE map $\hat{\mu}(r, \theta_{SiPM})$ (number of photoelectrons per unit of visible energy), which gives the expected light level to compare with the detected charge of the SiPMs to extract the radius and energy information. Due to the spherical symmetry of the detector, the nPE map is mainly related to the event radius \textbf{r} (=$|\Vec{r}|$) and the angle \text{$\theta_{SiPM}$} that the line joining the center and SiPM makes with the position vector $\Vec{r}$ of the event. Fig. \ref{event} represents the event parameters in the CD for an event at point ($\Vec{r}, \theta, \phi$). In Fig. \ref{templates}, nPE maps generated by simulating $^{68}\text{Ge}$ events at five different calibration points in the CD are shown. To calculate the nPE maps, the dark noise contribution due to SiPMs (which is considered as 20Hz/mm$^2$ in simulation \cite{Liu:2024cxo}) is removed (by subtracting the number of photoelectron contributions from dark noise), while cross-talk effect is included. The peaks around $\theta_{SiPM} =180^o$ are caused by the SiPM reflections. For the events that deposit energy at the points other than the calibration points, we use linear interpolation to calculate the nPE map for that point, by taking into consideration the nPE maps at adjoining calibration points. Other interpolation methods like Cubic and polynomial methods were also tried but they give similar results.

The nPE maps provide us the information of the expected number of photoelectrons $\hat{\mu}(r, \theta_{SiPM})$ per MeV detected by each SiPM for a given vertex. The probability of observing \textquotedblleft k\textquotedblright photoelectrons on an SiPM follows a Poisson distribution.  The probability of observed charge \textquotedblleft$q_i$\textquotedblright generated by the $i^{th}$ SiPM after being hit by \textquotedblleft k\textquotedblright  photoelectrons is given by P$^i(q_i|k)$ (which is considered gaussian). Given the observed and expected nPE information of SiPMs, a maximum likelihood method is developed to reconstruct the event vertex and energy simultaneously, using the likelihood function which is constructed as the product of:
 \begin{itemize}
       \item The probability of observing \textquotedblleft k\textquotedblright photoelectrons on $i^{th}$ SiPM which is defined as:
        \begin{equation}
       P^i(k,\mu_i) = \frac{\mu^{k}_{i}}{k!}e^{-\mu_i} 
       \end{equation} 
        \item The probability density of observed charge \textquotedblleft$q_i$\textquotedblright \\generated by the $i^{th}$ SiPM after being hit by \textquotedblleft k\textquotedblright \\ photoelectrons which is given as follows:
      \begin{equation}
       P^i(q_i|k) = \frac{1}{\sqrt{2\pi k}S_i}e^{\frac{-(q_i-kQ_i)^2}{2kS_i^2}} 
       \end{equation} 
       \begin{equation}
       \mu_i(r,E) = E \cdot \hat{\mu}_i(r,\theta_{SiPM})
       \label{expected-charge}
       \end{equation} 
\end{itemize}
The term $\mu_i(r,E)$ is the expected number of PE for $i^{th}$ SiPM which depends on its position relative to the event vertex(as shown in Fig.\ref{event}) as well as on the vertex and energy of the event, as defined in Eq. \ref{expected-charge} . $\hat{\mu}_i(r,\theta_{SiPM})$ is the the expected PE number per unit of visible energy for $i^{th}$ SiPM, and is provided by the nPE maps. An important point to remember is that $\theta_{SiPM}$ is not a free parameter. \text{$Q_i$} and \text{$S_i$} are the mean and sigma of the single photoelectron spectrum (SPES) of SiPM respectively. \text{E} is the energy calculated after each iteration of minimization of likelihood function. The probability of observing a hit pattern for all \enquote{n} channels  ($q_1$,$q_2$,$q_3$,......,$q_n$) for an event can be written as:
  \begin{equation}
       p( \{ q_i\} |r,E) = \prod_{unhit}e^{-\mu_j}\prod_{hit} \left( \sum_{k=1}^{\infty}P^i(k,\mu_i)\times P^i(q_i|k)\right)
       \label{total-function}
\end{equation}
 The likelihood function ($\mathcal{L}$) of Charge Maximum Likelihood Estimation (QMLE) method is constructed using Eq.\ref{total-function} and is given in Eq. \ref{likelihood}, which on minimization gives energy and radius using only charge information. During minimization, \textquotedblleft r\textquotedblright and \textquotedblleft$E$\textquotedblright are treated as free parameters whereas $\theta$ and $\phi$ are kept fixed to the values provided by the CCA \cite{Shi:2025zut} . The minimization requires an initial estimate of the event radius which is provided by the output of the CCA.

\begin{equation}
       \mathcal{L}(\{q_i\}|r,E) = -ln(p(\{q_i\}|r,E))
       \label{likelihood}
\end{equation}
It is general practice to take the logarithm which would replace continuous multiplications by continuous additions. Due to the monotonic behaviour of logarithmic functions, it’s more convenient to minimize -ln. When $\mathcal{L}$ reaches its minimum, the most probable value of r and E are derived as reconstructed radius ($R_{rec}$) and reconstructed energy ($E_{vis}^{rec}$).\\

\begin{figure}[!h]
	\includegraphics
	[width= 12cm, height=10cm]
	{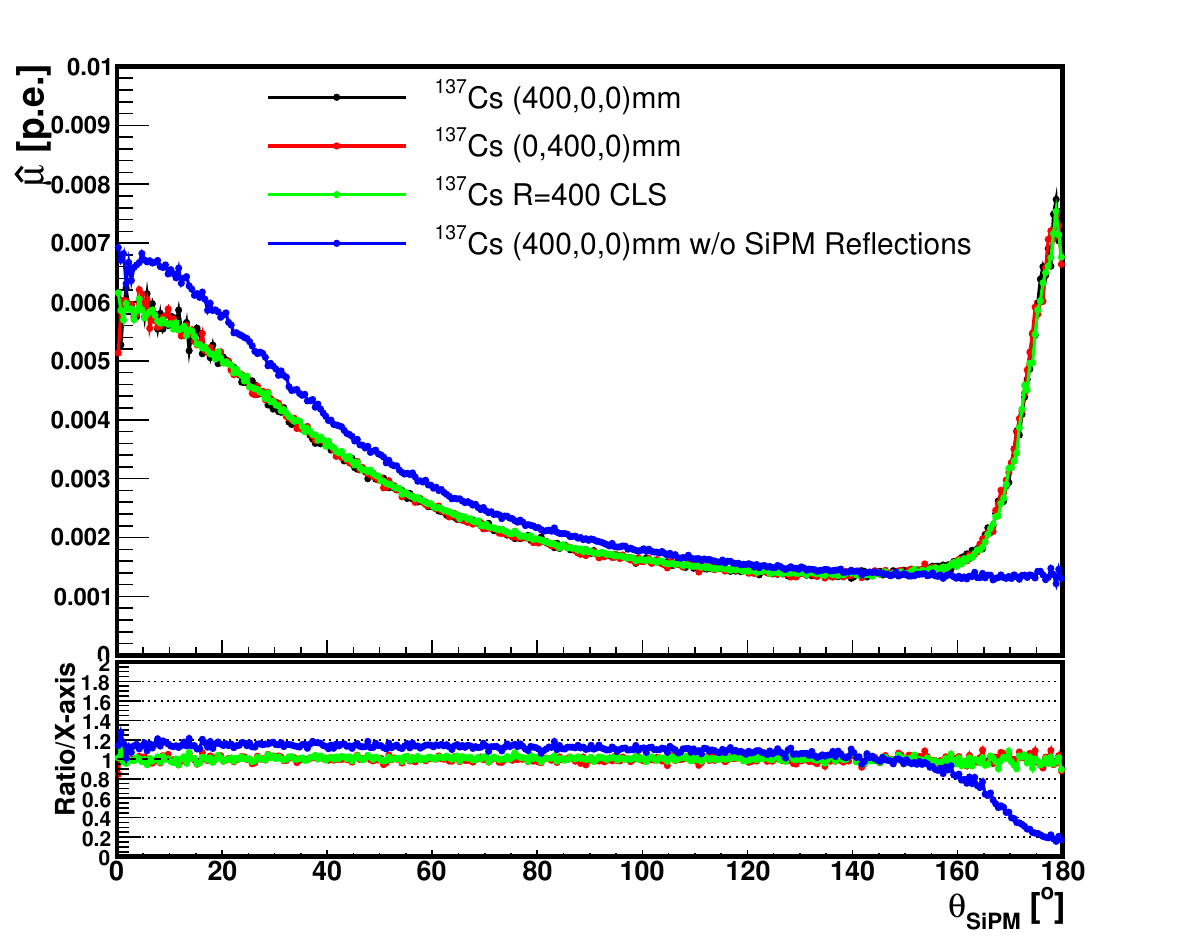}
	\caption{nPE maps generated by $^{137}\text{Cs}$ (0.662 MeV gamma source) at r = 400 mm along X-axis, Y-axis and CLS respectively. nPE map without SiPM reflection is also shown for comparison}
	\label{templates_Cs_diffaxes}
\end{figure}

As discussed earlier, the main ingredient for this algorithm is the charge templates as shown in Fig.  \ref{templates}. In case of real data, these charge templates will be generated by using the calibration data. In the detector, the calibration will be performed along the Z-axis by the ACU and in an off Z-axis (which is a plane) by the CLS. As the detector has a spherical symmetry, it is therefore important to check the distributions of the templates generated along different directions. Fig. \ref{templates_Cs_diffaxes} shows the comparison of the nPE maps generated by simulating the $^{137}\text{Cs}$ source at a radius of 400 mm from the center of the detector along X-axis, Y-axis and CLS respectively. From the figure it can be seen that the nPE maps are perfectly superimposed which is possible due to the spherical symmetry of the detector. As mentioned earlier, the rise of the distribution around $\theta_{SiPM} = 180^o$ is due to the SiPM reflections. In Fig. \ref{templates_Cs_diffaxes}, the blue curve shows the nPE map generated by the $^{137}\text{Cs}$ source at the point (400, 0, 0) mm without considering the SiPM reflections. As can be seen, reflections not only introduce a peak around the $\theta_{SiPM} = 180^o$ region but also shifts the nPE distribution a bit lower for the region $\theta_{SiPM} < 150^o$. The effect of the reflections on the overall reconstructed energy resolution will have to be considered. Similarly the overlapping of templates generated at a radius of 400 mm along different axes (black, red and green) shown in Fig. \ref{templates_Cs_diffaxes} shows that any set of nPE maps obtained from one axis can be used as an input to the reconstruction algorithm. Therefore, nPE maps can be completely defined along a single axis.

\begin{figure}[!h]
	\includegraphics
	[width= 12cm, height=9cm]
	{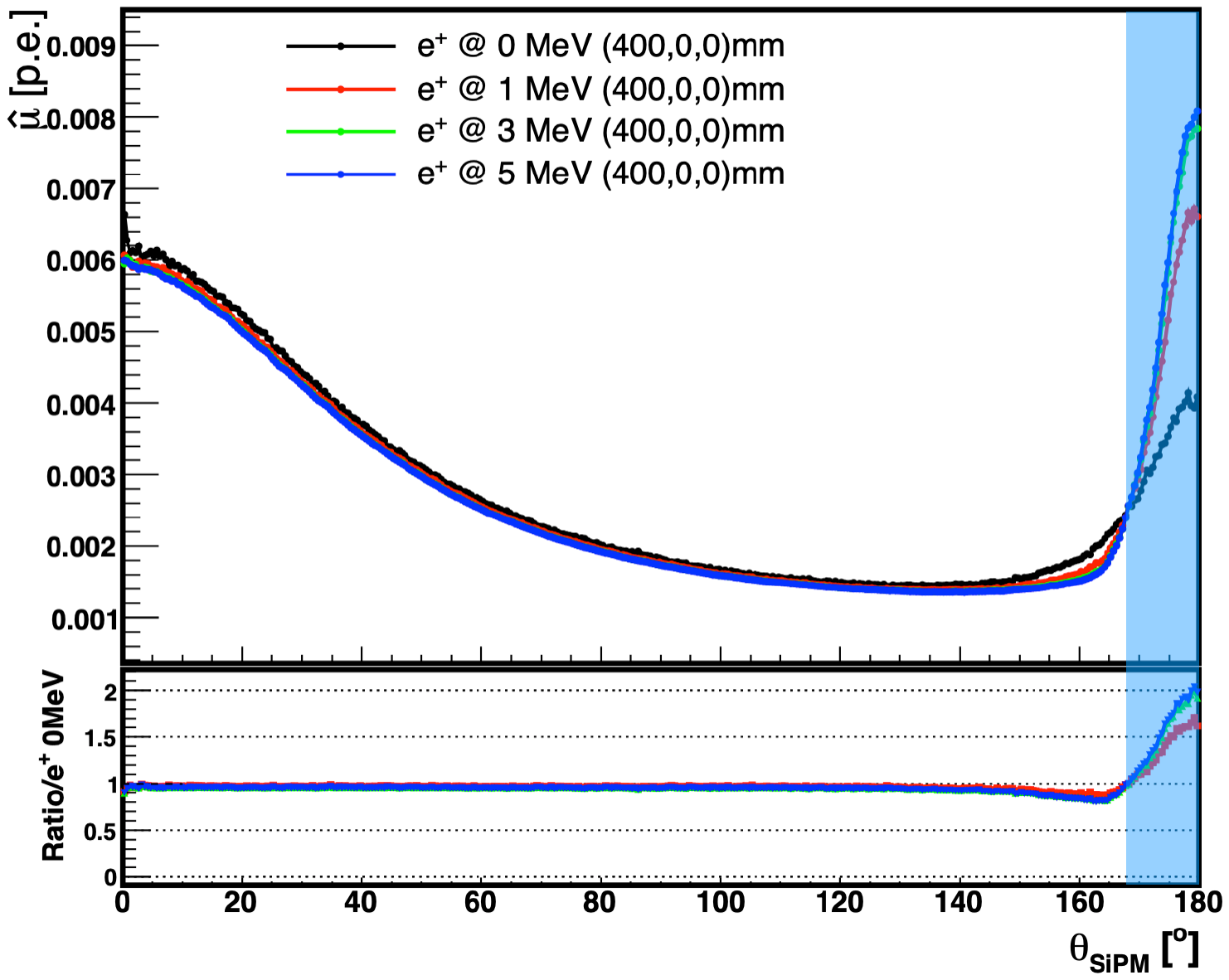}
	\caption{nPE maps generated at (400, 0, 0) mm by $e^{+}$ with kinetic energy of (0, 1, 3, 5) MeV respectively.}
	\label{templates_Ge_diffenergies}
\end{figure}

It is also need to be seen the comparison of the nPE maps at different energies. Fig. \ref{templates_Ge_diffenergies} shows the simulated nPE maps generated by $e^+$ at the point (400, 0, 0) mm with kinetic energies of (0, 1, 3, 5) MeV respectively. As can be seen from the figure, the peak region (shaded region) around $\theta_{SiPM} = 180^o$ shows an increasing trend with energy. This could be due to the fact that with increase in energy more photons are getting reflected to this region.

\begin{table}[!h]
	\caption{MC generated samples for prompt signal.}
	\label{positron-samples}
	\begin{tabular*}{8cm} {llll}
		\hline
		Source & Kinetic Energy (MeV) & Positions (cm) \\
		\hline
		$e^{+}$  & 0,1,2,3 & 0,10,20,30,40,50,60,70 \\
		\hline
		$e^{-}$  & 1,2,3,4 & 0,10,20,30,40,50,60,70 \\
		\hline
	\end{tabular*}
\end{table}

\section{Results}
\label{Results}
\subsection{Vertex Reconstruction}
\begin{figure}[h!]
	\centering     
	\subfigure[] {\label{Bias-positrona}\includegraphics[width=2.7in,height=2.4in]{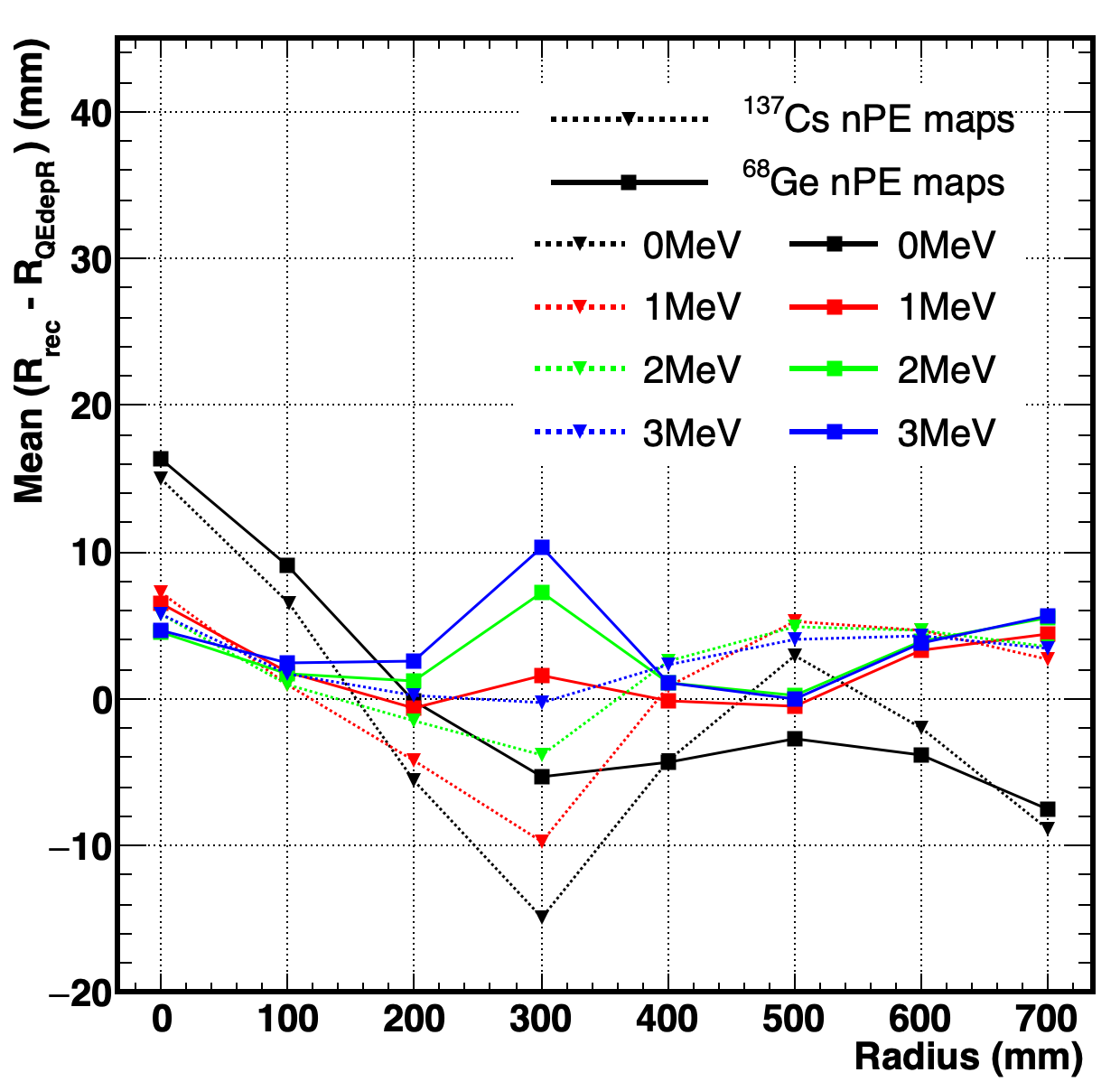}}
	\subfigure[] {\label{Bias-positronb}\includegraphics[width=2.7in,height=2.4in]{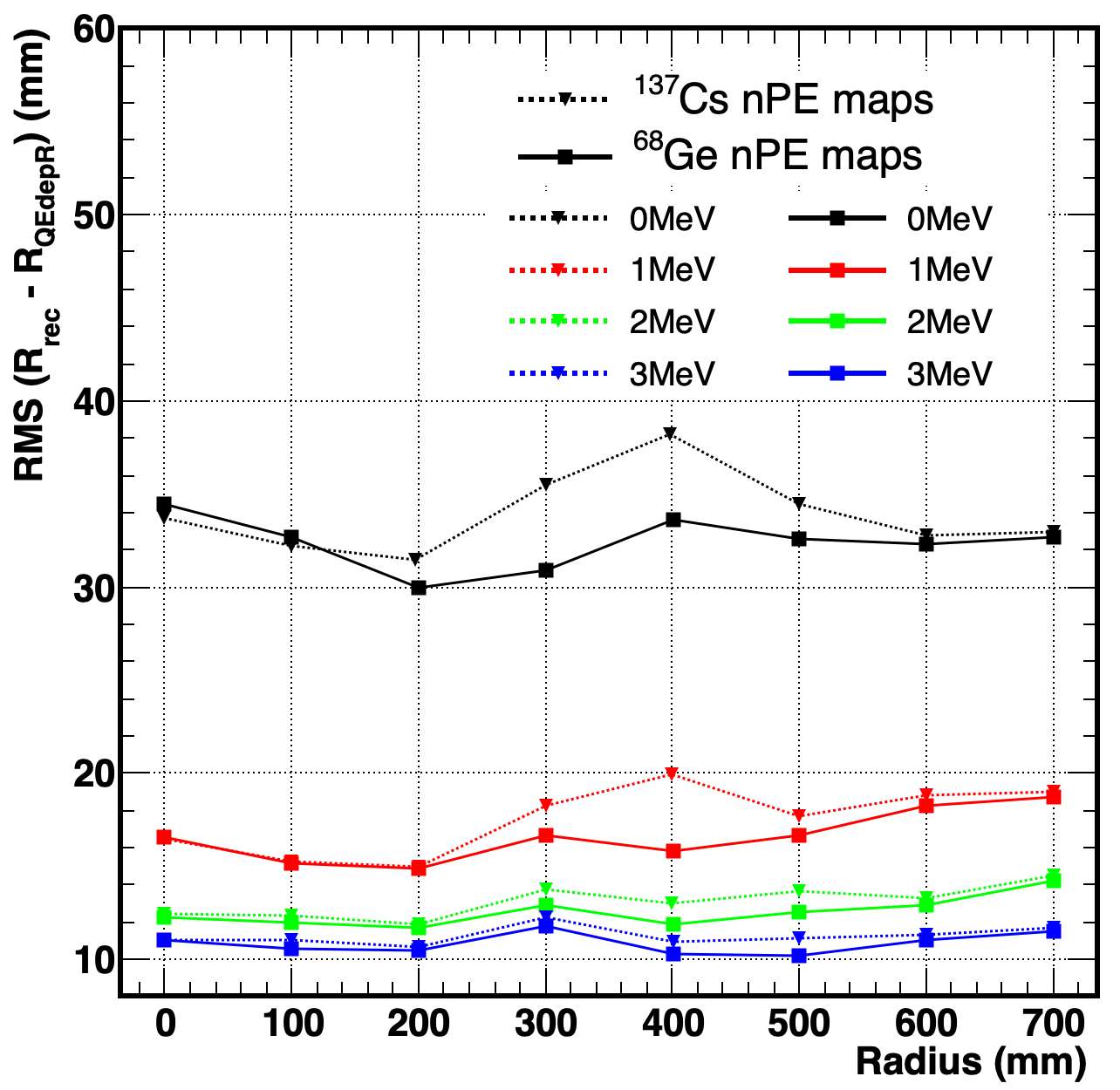}}
	\caption{Bias(left panel) and Resolution (right panel) of the reconstructed radius for $e^+$ at different energies.}
	\label{Biaspositron}
\end{figure}

\begin{figure}[h!]
	\centering     
	\subfigure[] {\label{Bias-electrona}\includegraphics[width=2.7in,height=2.4in]{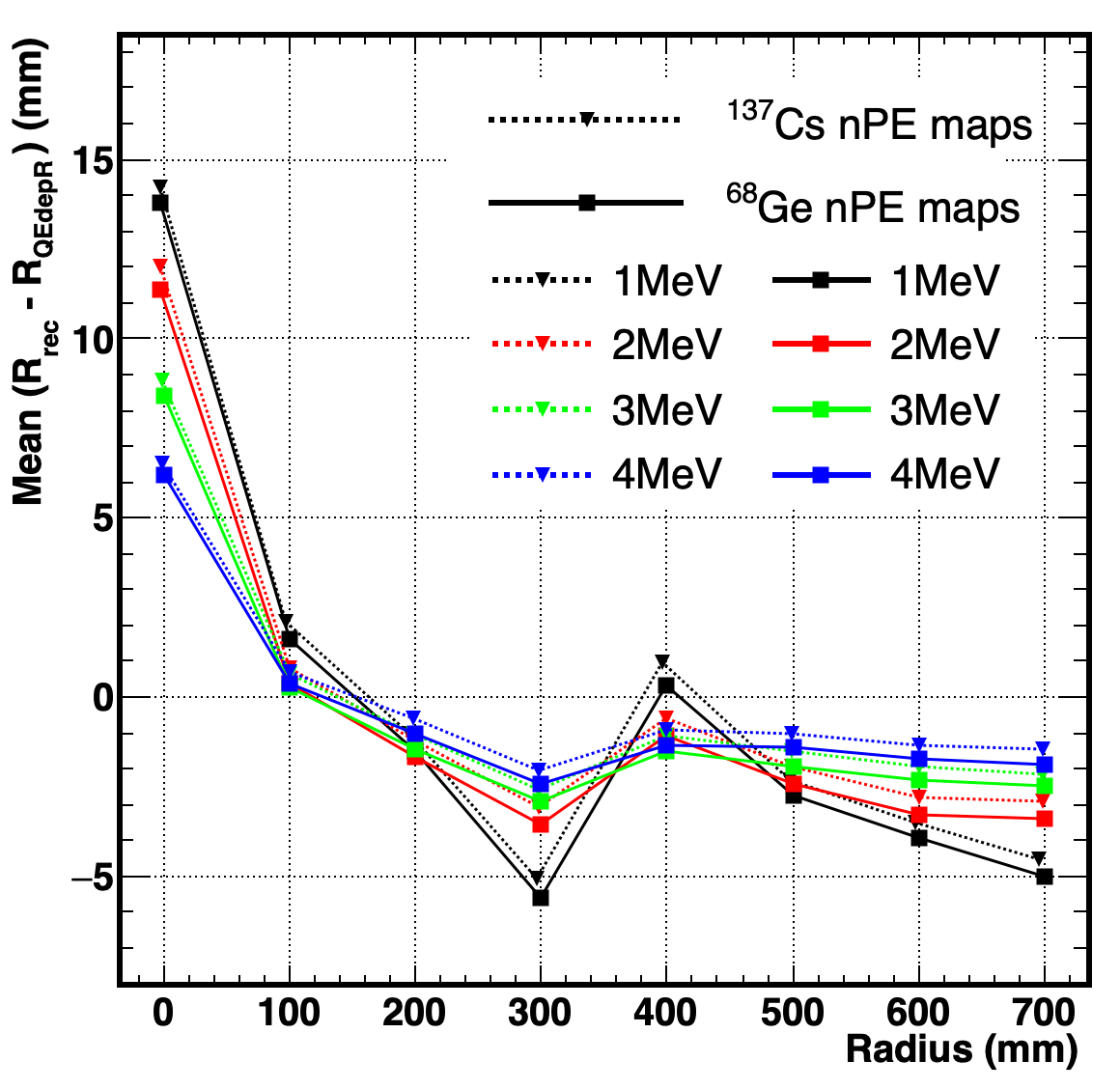}}
	\subfigure[] {\label{Bias-electronb}\includegraphics[width=2.7in,height=2.4in]{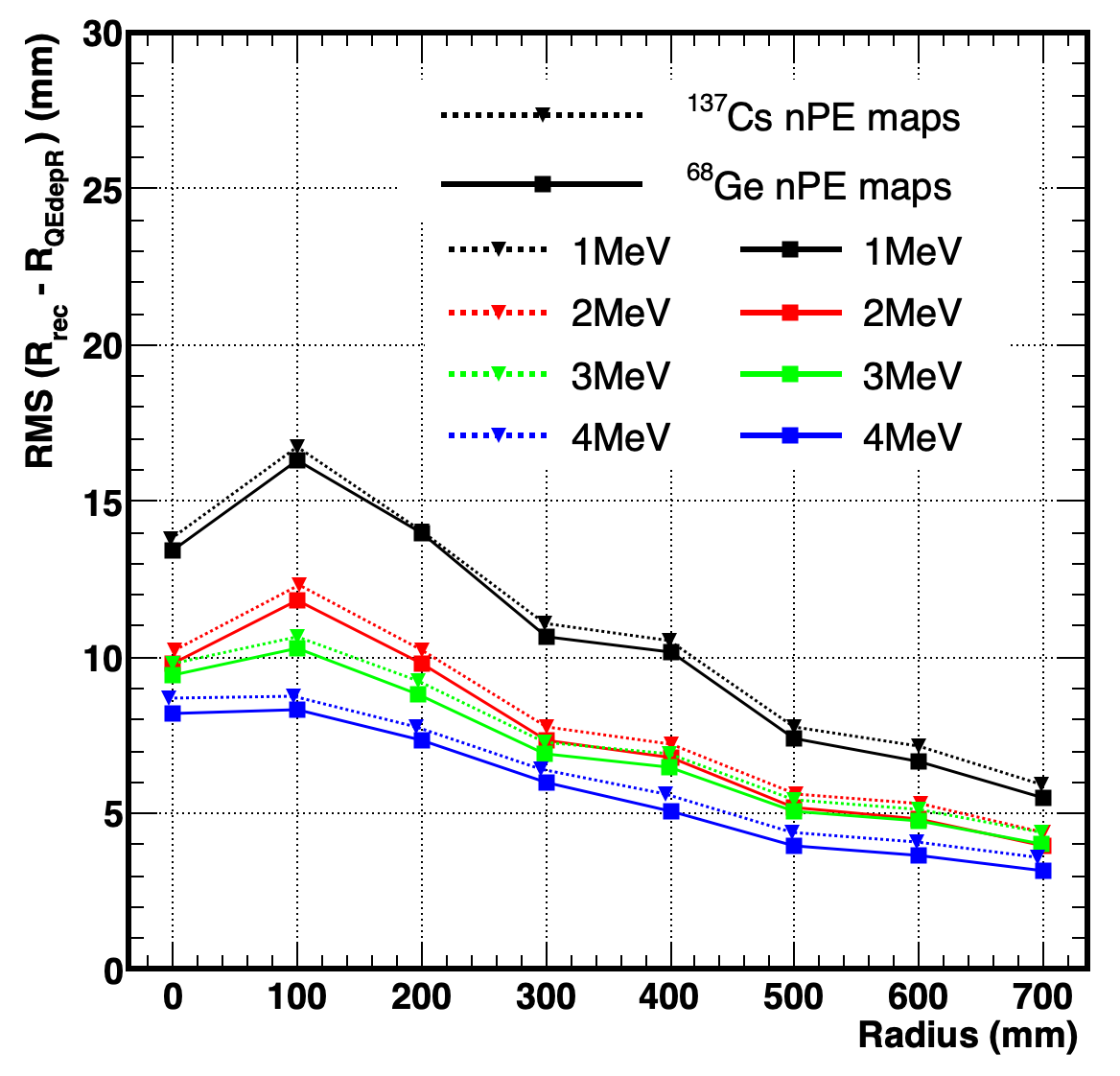}}
	\caption{Bias(left panel) and Resolution (right panel) of the reconstructed radius for $e^+$ at different energies.}
	\label{Bias-electron}
\end{figure}

To study the performance of the QMLE reconstruction method, Monte Carlo (MC) simulations were utilized. These simulations included thorough modeling of the detector, electronics, and calibration, allowing for a more realistic representation of the detector’s charge and timing responses. This served as the data source for the creation and assessment of the reconstruction algorithms. In order to reconstruct the prompt signal of IBD events, a set of positron samples were generated. The information of the positron samples used in this paper are summarized in Table \ref{positron-samples}. To generate all these samples, the detector simulation is done based on Geant4 \cite{GEANT4:2002zbu} detector simulation. All properties of LS \cite{JUNO:2020bcl} and details about the optical processes of photons propagating in LS are also implemented \cite{Zhang:2020mqz}. Real detector geometry such as the arrangement of the SiPMs as well as missing SiPMs (at north and south pole openings of CD) and the supporting structures are also considered in the simulation.

The reconstructed vertex is compared with the true vertex from the MC samples. The radius bias and radius resolution are defined as the mean and standard deviation of the \text{$R_{rec}-R_{QEdep}$} distribution respectively, where $R_{QEdep}$ is the radius of the energy-deposit center point after quenching in the MC simulation. The radial bias of the reconstruction is shown in Fig. \ref{Bias-positrona}, where different colors represent events with different energies. The reconstruction is performed by using the nPE maps generated by $^{137}\text{Cs}$ and $^{68}\text{Ge}$ sources. An energy-dependent bias behavior is noticeable at the center as well as near the detector boundary. At the center the radial bias decreases with energy but around the detector boundary, bias increases with increase in energy.

\begin{table}[htb]
\centering
\caption{Vertex bias and resolution at the center of CD for $e^+$ and $e^-$ samples with a kinetic energy of 0 MeV and 1 MeV respectively.}
\begin{tabular}{|c|c|c|c|c|}
\hline
\textbf{Particle} & \multicolumn{2}{|c|}{\textbf{Bias [mm]}} & \multicolumn{2}{|c|}{\textbf{Resolution [mm]}}\\
\cline{2-5}  
 & \textbf{$^{137}$Cs} & \textbf{$^{68}$Ge}& \textbf{$^{137}$Cs} & \textbf{$^{68}$Ge}\\
\hline
$e^+$ & 15 & 16 &34 &34.5 \\
\hline
$e^-$ & 14 & 14 &14 & 13.5 \\
\hline
\end{tabular}
\label{Bias-table}
\end{table}
The spatial resolution of the vertex reconstruction for $e^+$ as a function of radius for different energies is shown in Fig. \ref{Bias-positronb}. The black dotted lines and black solid lines show the radial resolution for $e^+$ with 0 MeV kinetic energy with $^{137}\text{Cs}$ and $^{68}\text{Ge}$ nPE maps respectively. As can be seen, the $^{68}\text{Ge}$ nPE maps give slightly better radial resolution than the $^{137}\text{Cs}$ nPE maps, whereas at higher energies, similar resolution is given by both types of nPE maps. This could be due to the fact that $^{68}\text{Ge}$ gives a signal
closer to $e^+$ than Cs, as it gives 2 gammas, so it is expected to give better results.\\

Fig.\ref{Bias-electrona} shows the vertex bias for the reconstruction of the $e^-$ events. Both type of nPE maps produce very similar results for similar energies. Fig.\ref{Bias-electronb} shows the vertex resolution for the reconstruction of the $e^-$ events. The radial resolution is much better for electron than positron because electron loses all its energy in the form of ionization and excitation and thus light comes from a compact region along the track. On the other hand positron loses its kinetic energy before it annihilates to produce two 0.511 MeV gammas. These gammas produce smearing (and hence extended light region) in the detector and leads to poor radial resolution. The radial resolution is similar for both type of nPE maps at all values of energy. Table \ref{Bias-table} summarizes the vertex bias and  resolution for $e^+$ and $e^-$ samples at the center of the detector at the total energy of 1.022 MeV and 1 MeV respectively.

\subsection{Energy Reconstruction}
As discussed earlier and shown in the Fig. \ref{templates_Ge_diffenergies}, the nPE maps in the reflection region show an energy dependence. To check the effect of this energy dependence on the reflection, reconstruction was performed for $e^+$ samples at the center and the edge of the fiducial volume (i.e radii 0 mm and 650 mm) by excluding the reflection region of the maps. This means that charge information from the SiPMs with $\theta_{SiPM}$ $<$ $160^o$ are considered for reconstruction. The relative change in the overall resolution at $E_k$ = 0 MeV was calculated to be less than \text{$1\%$}.\\

\begin{figure}[!h]
	\includegraphics[width=0.75\hsize]{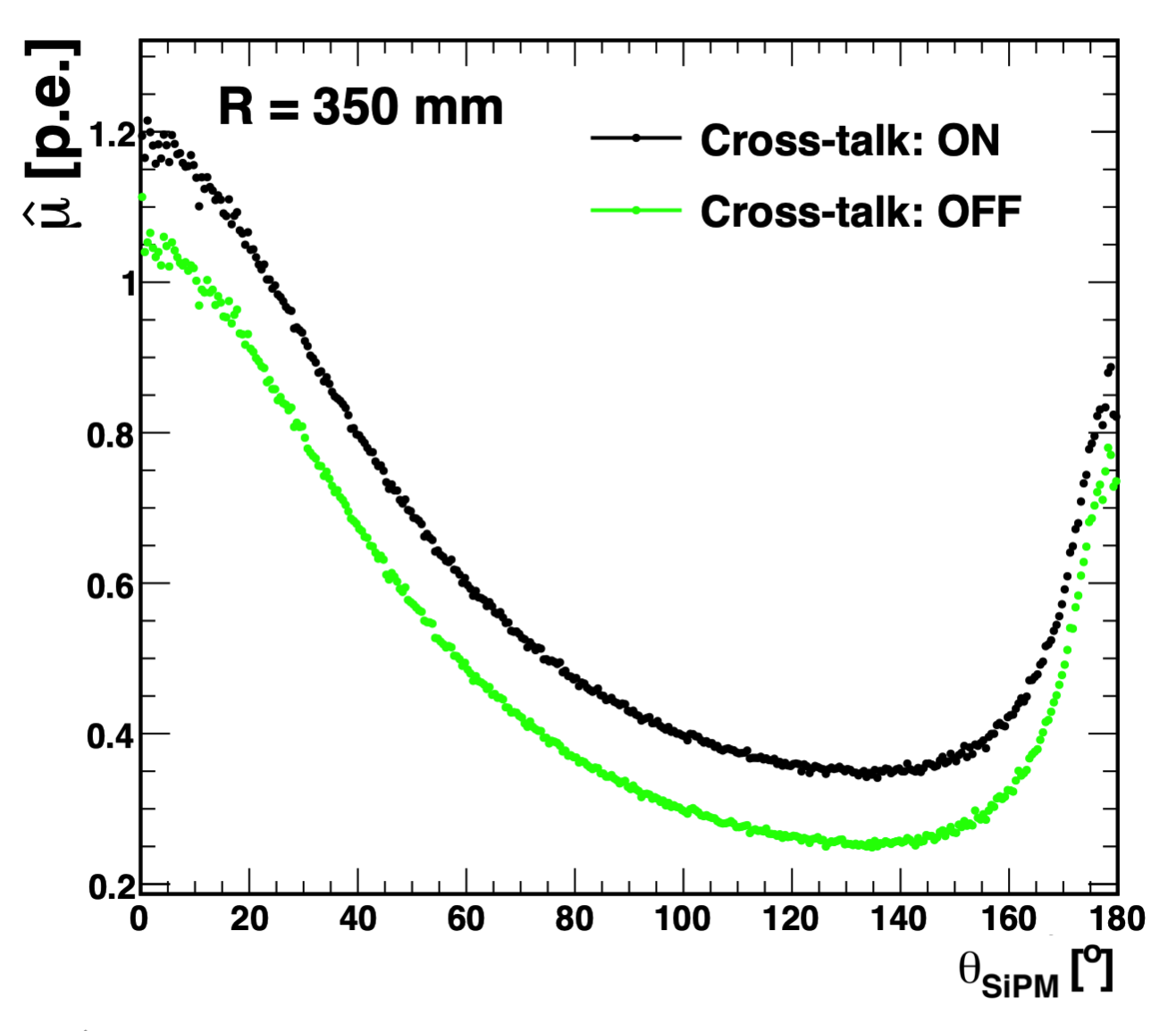}
	\caption{ Comparison of the nPE maps generated with and without considering cross-talk effect.}
	\label{templates-comparison-xtalk}
\end{figure}

Another factor that affect the energy resolution would be the cross-talk effect (internal and external cross-talk). In SiPMs, the term \enquote{internal cross-talk} refers to an effect wherein the avalanche breakdown of one microcell (or pixel) inadvertently causes the avalanching of neighboring microcells. \enquote{External cross-talk} (sometimes called inter-channel cross-talk) occurs between separate SiPMs or different channels in an array when light or electrical signals from one sensor trigger another nearby sensor. This results in additional, spurious signals that can complicate the interpretation of the actual event being measured. Cross-talk in SiPMs is a significant factor affecting their performance. To study the energy resolution degradation due to the cross-talk effect (both internal and external cross-talk), nPE maps were generated by disabling the cross-talk effect (internal and external) in the electronics simulation as shown in Fig. \ref{templates-comparison-xtalk}, and also the event samples were generated with this configuration. The reconstruction was performed for these event samples (without considering cross-talk) and the resolution was calculated at the center of the detector. The relative change to the energy resolution is around 0.5 $\%$. \\

In addition to the cross-talk effect, another crucial electronic effect is the dark noise of the SiPMs. Dark noise in SiPMs refers to the unwanted signals generated by the device in the absence of any incident light. This noise primarily originates from thermally induced carriers that trigger avalanche breakdown in the microcells, leading to output pulses similar to those caused by genuine photon detection events. It is an important consideration in applications where low-light levels are being detected, as it can affect the signal-to-noise ratio and overall performance of the SiPM. As discussed earlier, the dark noise is removed while calculating the nPE maps.The effect of dark noise on the overall resolution is also looked for by generating the  samples after switching off the dark noise (which is 20Hz/mm$^2$ in simulation). The resolution is calculated for these samples. The relative change to the overall energy resolution is around 0.16$\%$. 

An important detector effect that would affect the overall energy resolution is the liquid scintillator quenching. Due to this process some excited molecules release energy without radiation emission. As a result, the number of emitted scintillation photons is not linearly proportional to the deposited energy. There are various models used to describe this quenching effect, but the most common one is the one proposed by Birks \cite{Birks:1951boa,Chou:1952jqv}, as given by following Eq.,
\vspace{0.3cm}
\begin{equation}
\Delta E_q = \frac{\Delta E}{1+k_B\frac{dE}{dx}}
\label{Birks}
\end{equation}
\vspace{0.3cm}

where $k_B$ is Birks' constants (in MC simulation $k_B$ = 0.01205 g/cm$^2$/MeV \cite{JUNO:2024fdc}), $\frac{dE}{dx}$ is the stopping power, $\Delta E$ is the deposited energy before quenching, and $\Delta E_q$ is the quenched energy which is used to determine the mean value of the number of scintillation photons to be generated. For a fixed energy, the number of optical photons produced by scintillation follows a Poisson distribution due to the random nature of the molecular-scale energy transfer. However, the original Poisson distribution is distorted by fluctuations in the energy deposition due to quenching effects. The quenching effect is studied by generating templates and samples without quenching effect in the LS and the reconstruction is performed for these samples to calculate the energy resolution .\\ 
The nPE statistics is the most crucial factor for a very good energy resolution. Experiments like KamLAND \cite{KamLAND:2002uet}  and Borexino \cite{Redchuk:2020igv} reported PE yields of 511 nPE/MeV and 250 nPE/MeV, respectively, whereas JUNO \cite{JUNO:2025gmd} is expected to have nPE yield around 1785 nPE/MeV. Small detectors on the other hand, are expected to have this number around 4500 nPE/MeV\cite{JUNO:2020ijm}. This number is achievable due to various factors. The first one is the small size, which causes less photon absorption in the liquid scintillator. The second is the photo sensitive coverage. The third is the high photon detection efficiency. The fourth one is the low temperature of $-50^o C$ which increases the photon yield of LS\cite{Xia:2014cca}.\\
An electron deposits all of its energy as a point charge particle, therefore it has comparatively less contribution from the scintillation quenching effect. Small quenching effect would lead to production of more scintillation photons and photoelectrons. Due to this, the contribution of nPE statistics to the overall resolution is also smaller for electrons than positrons.  

\begin{figure}[!h]
	\includegraphics[width=0.65\hsize]{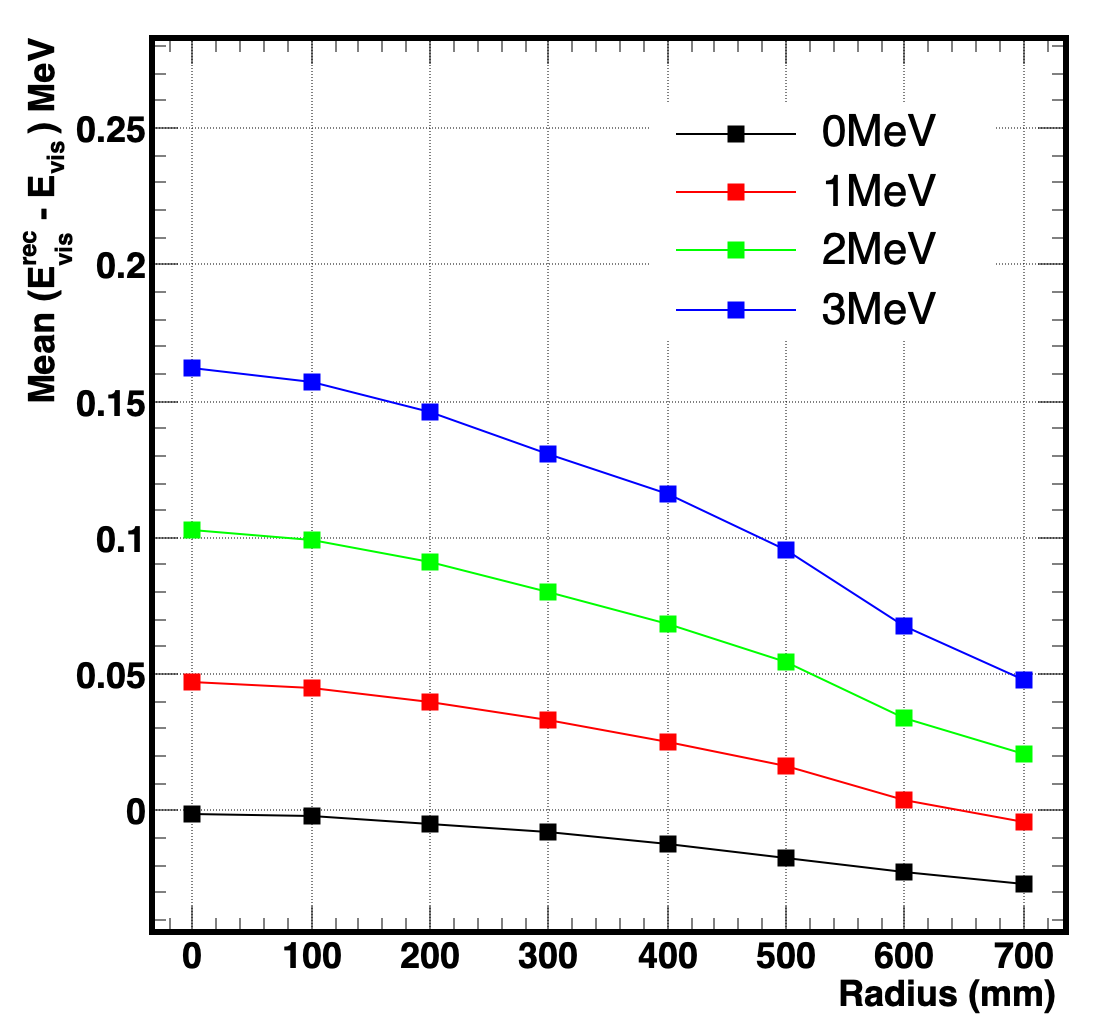}
	\caption{ Energy Bias for the reconstructed $e^+$.}
	\label{EB-positron}
\end{figure}

\begin{figure}[!hb]
	\includegraphics[width=0.65\hsize]{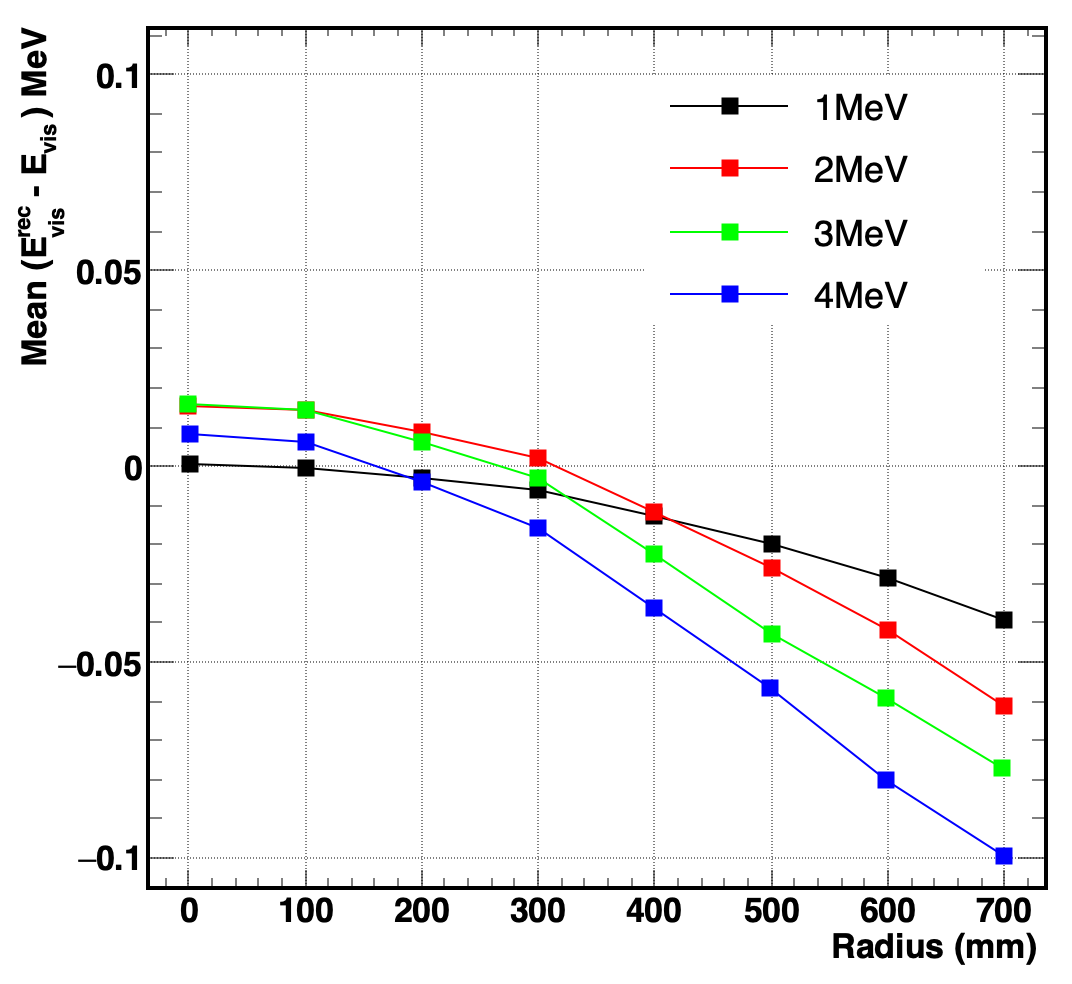}
	\caption{ Energy Bias for the reconstructed  $e^-$.}
	\label{EB-electron}
\end{figure}

Fig.\ref{EB-positron} shows the energy bias of the reconstruced $e^+$ at different energies. The figure shows that the bias increases as a function of energy. Similarly Fig.\ref{EB-electron} shows the energy bias for electron. In case of electron also, the energy bias increases with increasing energy. The energy dependence of the bias is attributed to the energy leakage which becomes prominent as energy increases.

\section{Conclusion}
\label{conclusion}
In this work we have demonstrated a method for the energy and radius reconstruction of the IBD events in the detector material using the charge information collected by the SiPMS. We detailed the data-driven methodology for constructing the nPE map, outlined the process for building the likelihood functions, and elucidated the reconstruction strategy. MC studies demonstrate that using this method, vertex bias less than 2 cm  and vertex resolution less than 4cm can be achieved for the positrons. As this method is template (nPE maps) dependent method, the template shapes show an energy dependent behaviour. The effect of this energy dependence of templates doesn't significantly impact the overall reconstructed energy and radial values. As for the energy resolution, the main contribution to resolution is the statistics of the number of photoelectrons (nPE). However, in addition to the statistical fluctuations in the detected nPE, several other effects impact the energy resolution in the detection of the IBD signals. The factors affecting the energy resolution are the scintillator quenching, cross-talk, SiPM reflections in addition to the contribution by photoelectron statistics. Apart from scintillator quenching which has significant contribution to overall energy resolution, the relative contribution of all other factors to the energy resolution is small.

\section{Acknowledgement}
We are gratefully for the JUNO-TAO and the technical staff of the participating institutions. This work was supported by the National Natural Science Foundation of China (Grant No. 12275281), the National Key Research and Development Program of China (Grant No. 2022YFA1602002). Special thanks Jiayang Xu for his support and Guihong Huang for his advice.




\end{document}